\documentclass[amsmath,twocolumn,superscriptaddress,prl,aps]{revtex4-1}
\usepackage{amsthm,amsfonts,graphicx,color,amsmath,amssymb,textcomp}
\usepackage{graphicx}
\usepackage{wasysym}
\usepackage{amsfonts}
\usepackage{bm}
\usepackage{enumerate}
\usepackage{color}
\usepackage{epstopdf}
\usepackage{latexsym}
\usepackage[breaklinks,colorlinks = true,linkcolor = red,urlcolor=cyan,citecolor=red]{hyperref}
\usepackage[caption=false,singlelinecheck=false]{subfig}

\usepackage[compat=1.1.0]{tikz-feynman}
\usepackage{times}
\usepackage{float}
\usepackage{changes}
\colorlet{Changes@Color}{black}

\usepackage{times}
\newcommand{\bea}{\begin{eqnarray}}
\newcommand{\eea}{\end{eqnarray}}
\newcommand{\be}{\begin{eqnarray}}
\newcommand{\ee}{\end{eqnarray}}
\newcommand{\bw}{\begin{widetext}}
\newcommand{\ew}{\end{widetext}}

\newcommand{\bs}{\boldsymbol}

\begin{document}
\title{Triplet-Superconductivity in Triple-Band Crossings}
\author{GiBaik Sim}
\email{gbsim1992@kaist.ac.kr}
\affiliation{Department of Physics, Korea Advanced Institute of Science and Technology, Daejeon 305-701, Korea}
\author{Moon Jip Park}
\email{moonjippark@kaist.ac.kr}
\affiliation{Department of Physics, Korea Advanced Institute of Science and Technology, Daejeon 305-701, Korea}
\author{SungBin Lee}
\email{sungbin@kaist.ac.kr}
\affiliation{Department of Physics, Korea Advanced Institute of Science and Technology, Daejeon 305-701, Korea}

\date{\today}
\begin{abstract}
Multi-band superconductivity in topological semimetals are the paradigms of unconventional superconductors. Their exotic gap structures and topological properties have fascinated searching for material realizations and applications. In this paper, we focus on triple point fermions, a new type of band crossings, and we claim that their superconductivity uniquely stabilizes spin-triplet pairing. Unlike conventional superconductors and other multi-band superconductors, such triplet superconductivity is the novel phenomena of triple point fermions where the spin-singlet pairing is strictly forbidden in the on-site interaction due to the Fermi statistics. We find
that two distinct triplet superconductors, characterized by the presence and absence of time-reversal symmetry,
are allowed which in principle can be controlled by tuning the chemical potential. For the triplet superconductor with time-reversal symmetry, we show that topologically protected nodal lines are realized. In contrast, for time-reversal broken case, the complication of topologically protected Bogoliubov Fermi surfaces emerges. Our theoretical study provides a new guidance for searching triplet superconductivities and their exotic implications.
\end{abstract}
\maketitle

The discovery of topological semimetallic phases have realized various types of new quasiparticles, characterized by topologically non-trivial band crossings. These quasiparticles are particularly interesting as its low-energy effective theory can mirror relativistic elementary particles. The representative examples are Dirac and Weyl semimetals mimicking the relativistic massless spin-$1/2$ fermions.\cite{burkov2011weyl,wan2011topological,young2012dirac,armitage2018weyl} They have gathered great interests due to the connection to the high-energy physics. Moreover, the condensed matter systems can realize even more exotic kinds of quasiparticle excitations that has no analogue of elementary particles in high energy physics.\cite{soluyanov2015type,muechler2016topological,wang2016hourglass,burkov2011topological,kim2015dirac,bzduvsek2016nodal,wieder2016double,bradlyn2016beyond,tang2017multiple,cano2019multifold,zhu2016triple}
Especially, recent studies show that the triple-band crossings are also realized at high-symmetry points\cite{bradlyn2016beyond,tang2017multiple} or on high-symmetry lines\cite{heikkila2015nexus,zhu2016triple,weng2016topological,weng2016coexistence,chang2017nexus,lv2017observation,ma2018three} stabilized by spatial symmetries. These quasiparticle excitations, referred to as triple point fermions, can carry the effective integer spin-$1$ since it is not constrained by the spin-statistics theorem. One representative low-energy theory, which captures triple point fermions, is characterized by linear band touching of two spin polarized bands with the Chern number $\pm 2$ and the existence of additional middle band with the trivial Chern number.\cite{bradlyn2016beyond} Such peculiar spin structures and energy dispersions of triple point fermions can have major impact on the nature of the correlated ground states in the presence of the many-body interactions. In particular, the possible unconventional superconducting states calls for concrete theoretical understanding.\cite{lin2018exotic}

The area of the unconventional superconductivity is characterized by the non-trivial pairing symmetries of the superconducting order parameters. Especially, it has recently been proposed that multi-band systems offer a new platform to achieve unconventional superconductivity.\cite{brydon2018bogoliubov,roy2019topological,szabo2018interacting,savary2017superconductivity,venderbos2018pairing,nomoto2016exotic,nomoto2016classification,yanase2016nonsymmorphic,kawakami2018topological,kim2018beyond} One example of such multi-band system is spin-orbit coupled $j\!=\!3/2$ system where possible realization of Cooper pair with total spin $S\!=\!2,3$ has been investigated.\cite{kim2018beyond,brydon2016pairing,agterberg2017bogoliubov,timm2017inflated,yang2016topological,yang2017majorana,yu2018singlet,kim2014spin,jeong2017direct} Moreover the system is known to offer generic routes to achieve pairing instabilities towards such unconventional superconductivity employing inter-band pairing channels.\cite{boettcher2018unconventional,sim2019topological} Despite growing interests in multi-band system, there have been few studies of superconductivity in $j\!=\!1$ system where Cooper pair with $S\!=\!0,2$ are forbidden to have even spatial parity by Fermi statistics.

In general, the spatial parity and the total spin of superconducting order parameter are not independent of each other. The Fermi statistics constrains them to be anti-symmetric under the exchange of two identical electrons forming a Cooper pair. For instance, the conventional $s$-wave BCS superconductors must be spin-singlet pairing, while the triplet superconductivity can only be realized with odd-parity order parameters. However, this scenario can be drastically changed and, indeed, inverted if we consider the pairing of pseudospin $j\!=\!1$ electrons. Specifically, the spin-singlet of the two composite spin $j\!=\!1$ fermions is symmetric under the exchange of the two spins. Accordingly, the formation of the spin-singlet pairs with even-parity is strictly forbidden, but the spin-triplet pairing is only allowed. This unique property of pseudospin $j\!=\!1$ electron motivates us the further investigations for the hunt of the new form of unconventional superconductivity.

In this work, we propose possible triplet superconducting ground states of triple point fermions. Using the Landau theory of the superconductivity, we discuss two distinct superconducting phases in the presence of SO(3) symmetry. They are characterized by distinct spin textures of the superconducting order parameters. Based on one-loop calculation, we find the time-reversal symmetric triplet pairing is energetically favored when the chemical potential lies far below the triple-band crossing point with the middle band having upward dispersion, which we refer to as `$(s_z)$' state. In this case, the triplet superconductors contain topologically protected nodal lines. On the other hand, with the chemical potential lying near or above the band crossing point, the middle band participates to pair with the other bands. This state breaks the time-reversal symmetry and the resulting spin texture of the order parameter resembles 3D chiral $p_x+ip_y$ superconductor, so we refer to it as `$(s_x\!+\!is_y)$' state. In this case, multiple Bogoliubov Fermi surfaces with finite Chern numbers emerge. Such unusual  triplet superconductor is the generic feature of triple point fermions and can be controlled by tuning the chemical potential.

To begin our discussion, we consider the $SO(3)$ and time-reversal symmetric low-energy theory of triple point fermions residing in two inequivalent valleys, which are time reversal partners.\cite{bradlyn2016beyond} Up to the quadratic order, the Hamiltonian can be expanded near the band crossing point as,
\bea
h_{\pm}(\bs k)\!=\!\psi_{\pm,\bs k} ^\dagger[(c|\bs{k}|^2-\mu)\mathbb{I}_3+v\boldsymbol{k} \cdot \boldsymbol{J} ]\psi_{\pm,\bs k}.
\label{eq:H}
\eea
Here, we define the three spinor as $\psi_{\pm,\bs k}\!=\!(\psi_{\pm,\bs k,1},\psi_{\pm,\bs k,0},\psi_{\pm,\bs k,-1})$. 
The first subscript $\pm$ and the second subscript $\pm1$ and 0 indicate the valley and the spin degree of freedom respectively.  $\bs J\!=\!(J_x,J_y,J_z)$ represents the $j\!=\!1$ angular momentum matrices. They are explicitly written as,
\begin{gather}
J_x \!=\! \frac{1}{\sqrt{2}}\left(
\begin{array}{ccc}
	0 & 1 & 0 \\
	1 & 0 & 1 \\
	0 & 1 & 0 \\
\end{array}
\right)
,
 J_y \!=\!\frac{1}{\sqrt{2}}\left(
\begin{array}{ccc}
	0 & -i & 0 \\
	i & 0 & -i \\
	0 & i & 0 \\
\end{array}
\right)
,
\nonumber
\\
J_z \!=\!
\left(
\begin{array}{ccc}
	1 & 0 & 0 \\
	0 & 0 & 0 \\
	0 & 0 & -1 \\
\end{array}
\right).
\end{gather}
$\mathbb{I}_3$ is the three-dimensional identity matrix. $v$ and $\mu$ are the effective linear velocity of the band crossings and the chemical potential respectively. $v\boldsymbol{k} \cdot \boldsymbol{J}$ term breaks inversion symmetry. $c|{\bs{k}}|^2$ term represents possible bending of bands and we assume $c>0$ without loss of generality. The Hamiltonian in Eq. \eqref{eq:H} has two bands with the dispersion, $\epsilon^{\pm1}(\bs{k})\!=\!\pm v|\bs{k}|+c|\bs{k}|^2-\mu$, having opposite spins, and the middle band with the dispersion, $\epsilon^{0}(\bs{k})\!=\!c|\bs{k}|^2-\mu$. These three bands can be characterized by the monopole charge of the Berry curvature, $C^{\pm1}\!=\!\mp2$ and $C^0\!=\!0$ respectively.

Prior to the description of the microscopic interactions, we first discuss the generic form of the allowed pairing order parameters. The pairing order parameters can be written as the sum of bilinear form, $\langle \psi_{+,\bs{k}}^\dagger g(\bs{K}\!+\!\bs{k})M_{S}\gamma \psi_{-,-\bs{k}}^* \rangle$, where the function $g(\bs{K}\!+\!\bs{k})$ describes the orbital part of the Cooper pair and $\gamma\!=\!e^{-i\pi J_y}$ is the unitary part of the time-reversal operator $\mathcal{T}=\gamma \mathcal{K}$ ($\mathcal{K}$ is the complex conjugate operator). Here, $\bs{K}$ indicates the position of the valley $+$ in momentum space. The matrix $M_{S}$ specifies the total spin, $S$, of the Cooper pair and is listed in Table \ref{tab:spin}. Composite of two $j\!=\!1$ spins can generate total spins up to spin-singlet ($S\!=\!0$), triplet ($S\!=\!1$) and quintet ($S\!=\!2$). Fermi statistics forces the order parameter to satisfy $g(-\bs{K}\!-\!\bs{k})(M_{S}\gamma)^T\!=\!-g(\bs{K}\!+\!\bs{k})M_{S}\gamma$. According to this condition, the even-parity pairings ($g(-\bs{K}\!-\!\bs{k})\!=\!g(\bs{K}\!+\!\bs{k})$) only allow spin-triplet pairing while the odd-parity pairings ($g(-\bs{K}\!-\!\bs{k})\!=\!-g(\bs{K}\!+\!\bs{k})$) allow spin-singlet and quintet pairings. Such combinations of the spatial parity and the spin shows exactly the opposite pattern from superconductors which are comprised of spin half-integer electrons. This is the key observation of our work. As a result, the superconducting state driven by the on-site interactions must be spin-triplet pairing state. In addition, it is worthwhile to note that this behavior is different from the spin polarized superconductors described by an effective $j\!=\!0$. For spin polarized case, the system requires dominant further-neighbor interactions to induce the triplet pairing, which is clearly distinct from our case.

\begin{table}[t!]
	\centering
	\begin{tabular}[c]{ccc}
		\hline\hline
		$S$ & $M_S$ & Spatial Parity \\
		\hline
		$0$ & $\mathbb{I}_3$ & Odd\\
		$1$ & $\sqrt{\frac32}(J_x,J_y,J_z)$ & Even \\
		$2$ & $(\Gamma_{x^2-y^2},\Gamma_{3z^2-r^2},\Gamma_{yz},\Gamma_{zx},\Gamma_{xy})$ & Odd\\
		\hline\hline
	\end{tabular}
	\caption{List of spin pairing matrices $M_S$. Electrons with $j\!=\!1$ can form a Cooper pair with total spin $S\!=\!0,1,2$. A Cooper pair with total spin $S$ is created by the operator $\psi_{+,\bs{k}}^\dagger g(\bs{K}\!+\!\bs{k})M_{S}\gamma \psi_{-,-\bs{k}}^*$. $M_S$ is normalized such that $\rm{tr}$$[M_S (M_S)^\dagger]\!=\!3$. Fermi statistics allow only the $M_{S=1}$ locally (momentum independent). While $M_{S=0,2}$ should be additionally multiplied by an odd power of momentum to satisfy Fermi statistics. The column ``Even,Odd'' indicates that the spatial parity of the superconducting order parameter. $\Gamma$ matrices are written as $\Gamma_{x^2-y^2}\!=\!\sqrt{\frac32}(J_x^2-J_y^2),\Gamma_{3z^2-r^2}\!=\!\sqrt{\frac12}(2J_z^2-J_x^2-J_y^2),\Gamma_{yz}\!=\!\sqrt{\frac32}(J_yJ_z+J_zJ_y),\Gamma_{zx}\!=\!\sqrt{\frac32}(J_zJ_x+J_xJ_z)$, and $\Gamma_{xy}\!=\!\sqrt{\frac32}(J_xJ_y+J_yJ_x)$. See main text for more details.}
	\label{tab:spin}
\end{table}

Motivated by the above discussions, we now consider the following form of the generic on-site interactions,
\bea
h_{int}\!=\!g\sum_{a=x,y,z}(\psi_{\bs{r}}^\dagger J_a \psi_{\bs{r}})^2.
\label{eq:int}
\eea
The interaction terms constitute a complete set of on-site interactions with SO(3) symmetry\cite{herbut2019role} and correspond to the interactions between $p$-wave-orbital densities since $\psi_{\bs{r}}^\dagger J_a \psi_{\bs{r}}$ transforms as $p$-wave orbitals. 
Here, we consider repulsive on-site interactions, $g\!>\!0$. To rewrite the interaction into pairing channels, we use the Fierz identity for electrons with pseudospin $j\!=\!1$.\cite{herbut2019role,herbut2009theory,vafek2010interacting,herbut2014topological,boettcher2018unconventional} The particle-hole channel interactions in Eq. \eqref{eq:int} are exactly rewritten into the form of the pairing channels as following,

\bea
h_{int}\!=\!-\frac{g}{2}\sum_{a=x,y,z}(\psi_{\bs{r}}^\dagger J_a\gamma \psi_{\bs{r}}^*)(\psi_{\bs{r}}^T (J_a\gamma)^\dagger \psi_{\bs{r}}). 
\label{eq:ph-pp}
\eea
We now find that there exists superconducting instability even when the on-site interactions are all repulsive ($g>0$). Based on the pairing interaction in Eq. \eqref{eq:ph-pp}, we now derive the Ginzburg-Landau(GL) free energy, $F(\vec{\Delta},T,v,\mu,g)$, as a function of the order parameter, $\vec{\Delta}\!=\!(\Delta_{x},\Delta_y,\Delta_z)$, where $\Delta_a\!=\!\langle \psi_{\bs{r}}^T (J_a\gamma)^\dagger \psi_{\bs{r}} \rangle$ corresponds to the $s$-wave spin-triplet pair with total spin $S\!=\!1$. By integrating out the electronic degrees of freedom, the free energy functional is written as,
\bea
\nonumber
F&\!=\!&r(T,v,\mu,g)|\vec\Delta|^2+q_1(T,v,\mu)|\vec\Delta|^4\\
&+&q_2(T,v,\mu)\sum_{a=1}^{3}|\vec\Delta^* \mathcal{I}_a \vec\Delta|^2
\label{eq:free}
\eea
where the matrix elements of $\mathcal{I}_a$ is given as $(\mathcal{I}_a)_{bc}\!=\!i\epsilon_{abc}$ where $\epsilon_{abc}$ is the Levi-Civita symbol.\cite{ho1998spinor,venderbos2018pairing} We find that Eq. \eqref{eq:free} can have the two possible superconducting ground states solely depending on the value of the coefficient $q_2$. When $q_2>0$, time-reversal symmetric state  with the order parameter $\vec\Delta\!=\!(0,0,1)$ is stabilized. For $q_2<0$, however, time-reversal broken state is stabilized with the order parameter, $\vec\Delta\!=\!(1,i,0)$. We note that the complex order parameters $\vec\Delta$ transformed under $SO(3)$ rotation are physically equivalent to the ones mentioned above. From now on, we call $\vec\Delta\!=\!(0,0,1)$ and $(1,i,0)$ states as $(s_z)$ and $(s_x\!+\!i s_y)$ states respectively. 

The $(s_z)$ and $(s_x\!+\!is_y)$ order parameters have distinct spin textures. The $M_S$ matrix of $(s_z)$ state is explicitly given as,
\bea
J_z\gamma \!=\! \begin{pmatrix} 0  & 0 & 1 \\ 0& 0& 0 \\ -1& 0& 0 \end{pmatrix}.
\eea
From the explicit form of the above matrix, we can observe that the $(s_z)$ state forms Cooper pairs with opposite spin components, using inter-band pairing. On the other hand, the $M_S$ matrix of $(s_x\!+\!is_y)$ state is given as,
\bea
\frac{(J_x+iJ_y)\gamma}{\sqrt2} \!=\! \begin{pmatrix} 0  & -1 & 0 \\ 1& 0& 0 \\ 0& 0& 0 \end{pmatrix}.
\eea
For $(s_x\!+\!is_y)$ state, if we consider an electron with $j_z\!=\!1$, it makes inter-band pairing with a $j_z\!=\!0$ electron. Similarly, $(s_x-is_y)$ state pairs a $j_z\!=\!-1$ electron with a $j_z\!=\!0$ electron.

By explicitly investigating the sign of $q_2$ within the leading one-loop calculation, one can determine the energetically favored state (See Supplementary Material for details). Fig.\ref{fig:pd} shows the calculated phase diagram as a function of dimensionless parameters, $\mu/T$ and $v/\sqrt T$, while keeping a dimensionless parameter $c\!=\!1/10$ with the momentum cutoff $\Lambda\!=\!\mu/v$. First of all, when the chemical potential lies far below the band crossing point, we find that the $(s_z)$ state is favored preserving time-reversal symmetry. However, when the chemical potential approaches to the band crossing point, the contribution of the middle band to the free energy become significant. We find that $(s_x\!+\!is_y)$ pairing is stabilized with the chemical potential lying near or above the band crossing point. In the limit where the middle band is perfectly flat ($c\!=\!0$), our one-loop calculation with momentum cutoff $\Lambda\!=\!\mu/v$ shows that $q_2$ is always negative, favoring $(s_x\!+\!is_y)$ state (See Supplementary Material for details).
\begin{figure}[t]
	\includegraphics[width=1\columnwidth]{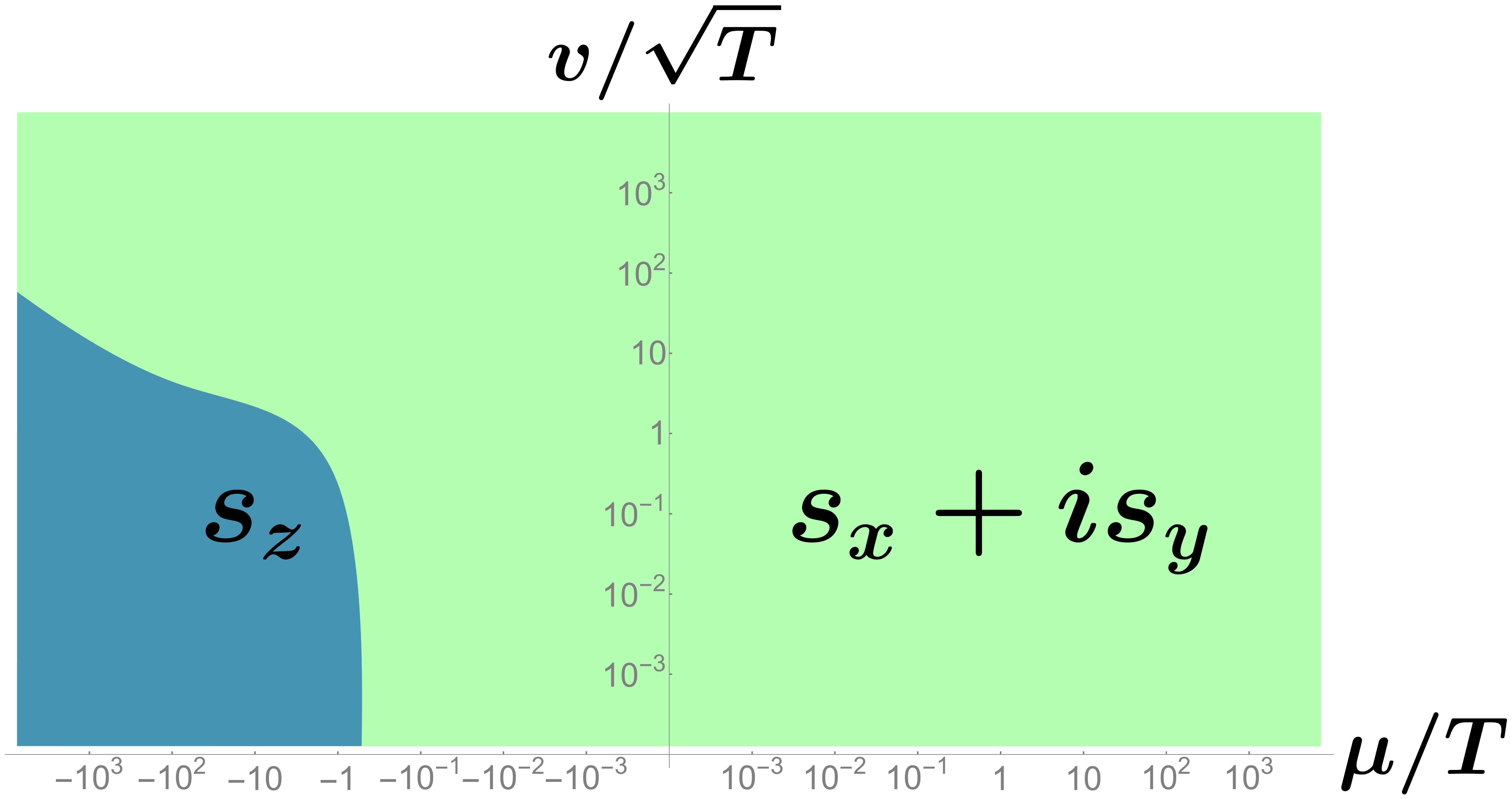}
	\caption{(color online) Phase diagram of topological superconductors with $s$-wave spin-triplet pair, as a function of $\mu/T$ and $v/\sqrt T$ with $c\!=\!1/10$. We find two distinct states: time-reversal symmetry broken state with $(s_x\!+\!is_y)$ pairing (green) and time-reversal symmetric state with $(s_z)$ pairing (blue). See main text for more details.}
	\label{fig:pd}
\end{figure}

After constructing the Landau theory and phase diagram  of the superconductivity, we now discuss the  Bogoliubov-de Gennes (BdG) quasiparticle spectrum of the superconducting states. The BdG Hamiltonian reads $\sum_{\boldsymbol{k}} \Psi^\dagger_{\boldsymbol{k}} H_{BdG}(\bs{k})\Psi_{\boldsymbol{k}}$ with 
\bea
H_{BdG}(\bs{k})\!=\!\left(
\begin{array}{cc}
	\hat{h}(\boldsymbol{k}) & \hat{\Delta} \\
	\hat{\Delta}^{\dagger} & -\hat{h}^{T}(-\boldsymbol{k}) \\
\end{array}\right)
\label{eq:bdg}
\eea
where $\Psi^\dagger_{\boldsymbol{k}}\!=\!(\psi_{+,\bs k}^\dagger, \psi_{-,-\bs k})$ and $\hat{h}(\boldsymbol{k})\!=\!(ck^2-\mu)\mathbb{I}_3+v\boldsymbol{k} \cdot \boldsymbol{J}$. Here, $\hat{\Delta}\!=\!|\Delta|J_z\gamma$ for the $(s_z)$ state and $\hat{\Delta}\!=\!|\Delta|(J_x+iJ_y)\gamma$ for the $(s_x\!+\!is_y)$ state where $\Delta$ is a real constant. 
\begin{figure*}[t!]
		\begin{minipage}{.245\linewidth}
			\centering
			\subfloat[$(s_z)$ state with $\mu\!<\!0$]{\label{fig:main_a}\includegraphics[scale=0.35]{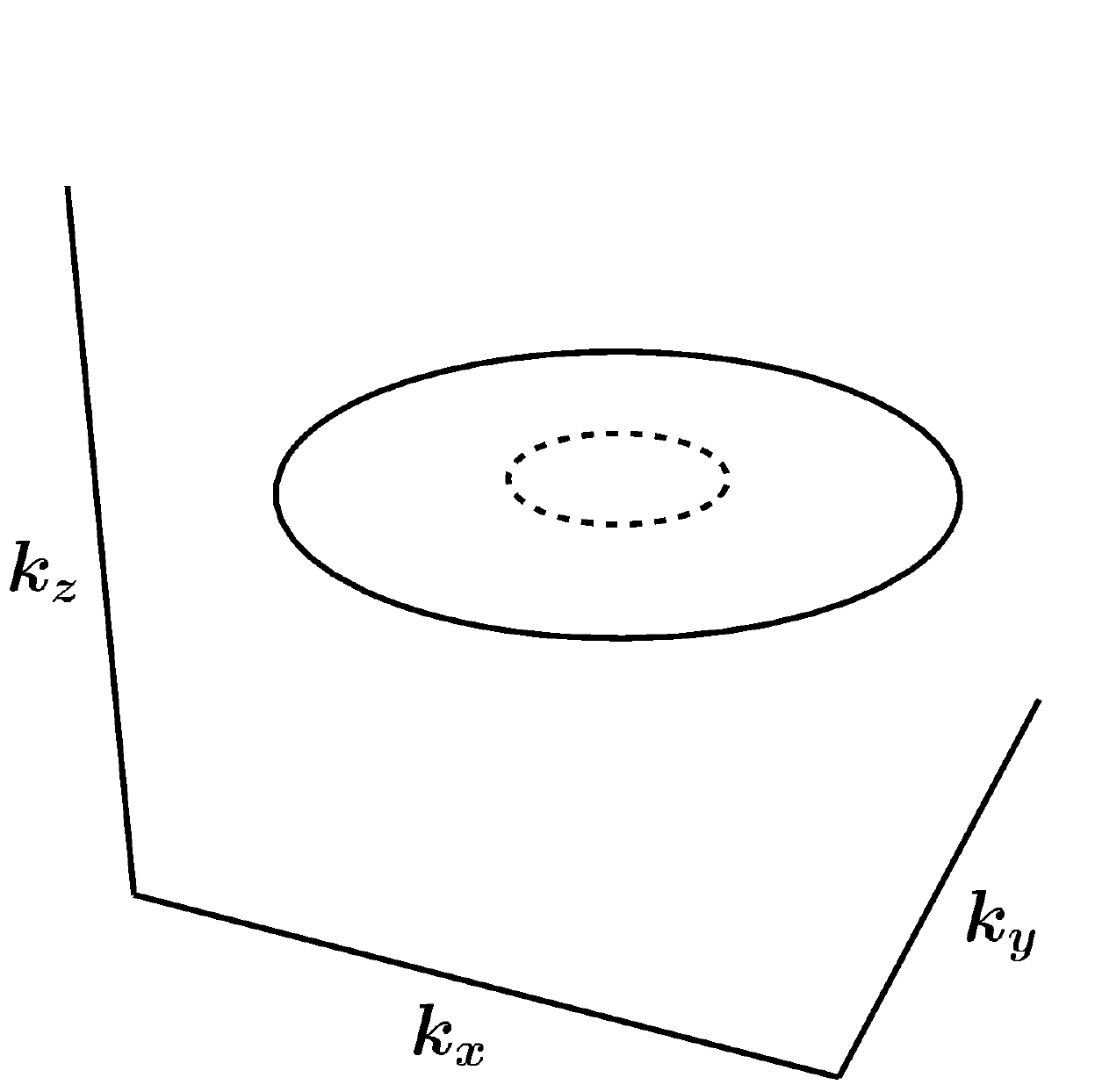}}
		\end{minipage}%
		\begin{minipage}{.245\linewidth}
			\centering
			\subfloat[$(s_x\!+\!is_y)$ state with $\mu\!<\!0$]{\label{fig:main_b}\includegraphics[scale=0.35]{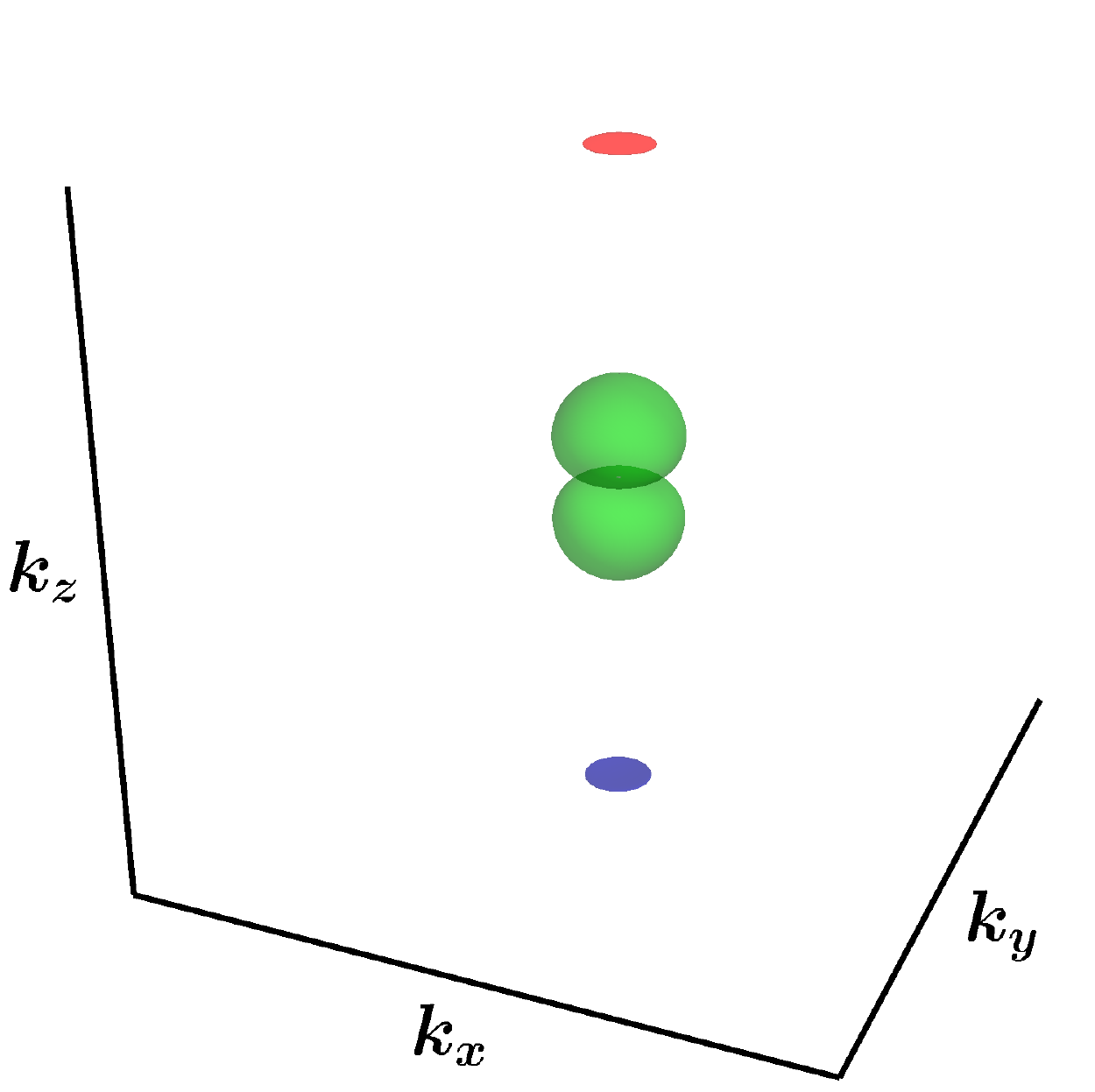}}
		\end{minipage}%
		\begin{minipage}{.245\linewidth}
			\centering
			\subfloat[$(s_x\!+\!is_y)$ state with $\mu=0$]{\label{fig:main_c}\includegraphics[scale=0.35]{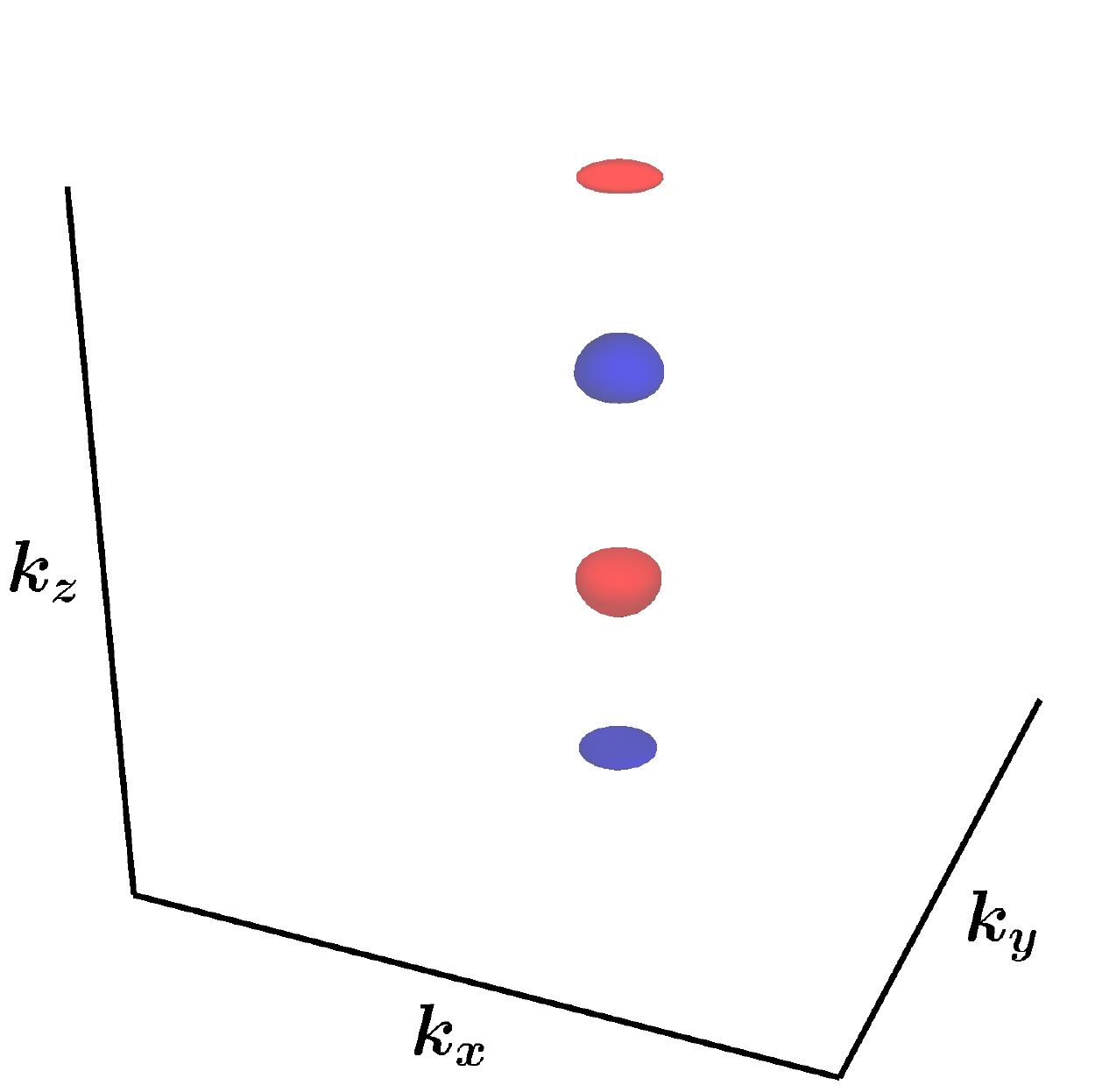}}
		\end{minipage}
		\begin{minipage}{.245\linewidth}
		    \centering
		    \subfloat[$(s_x\!+\!is_y)$ state with $\mu\!>\!0$]{\label{fig:main_d}\includegraphics[scale=0.35]{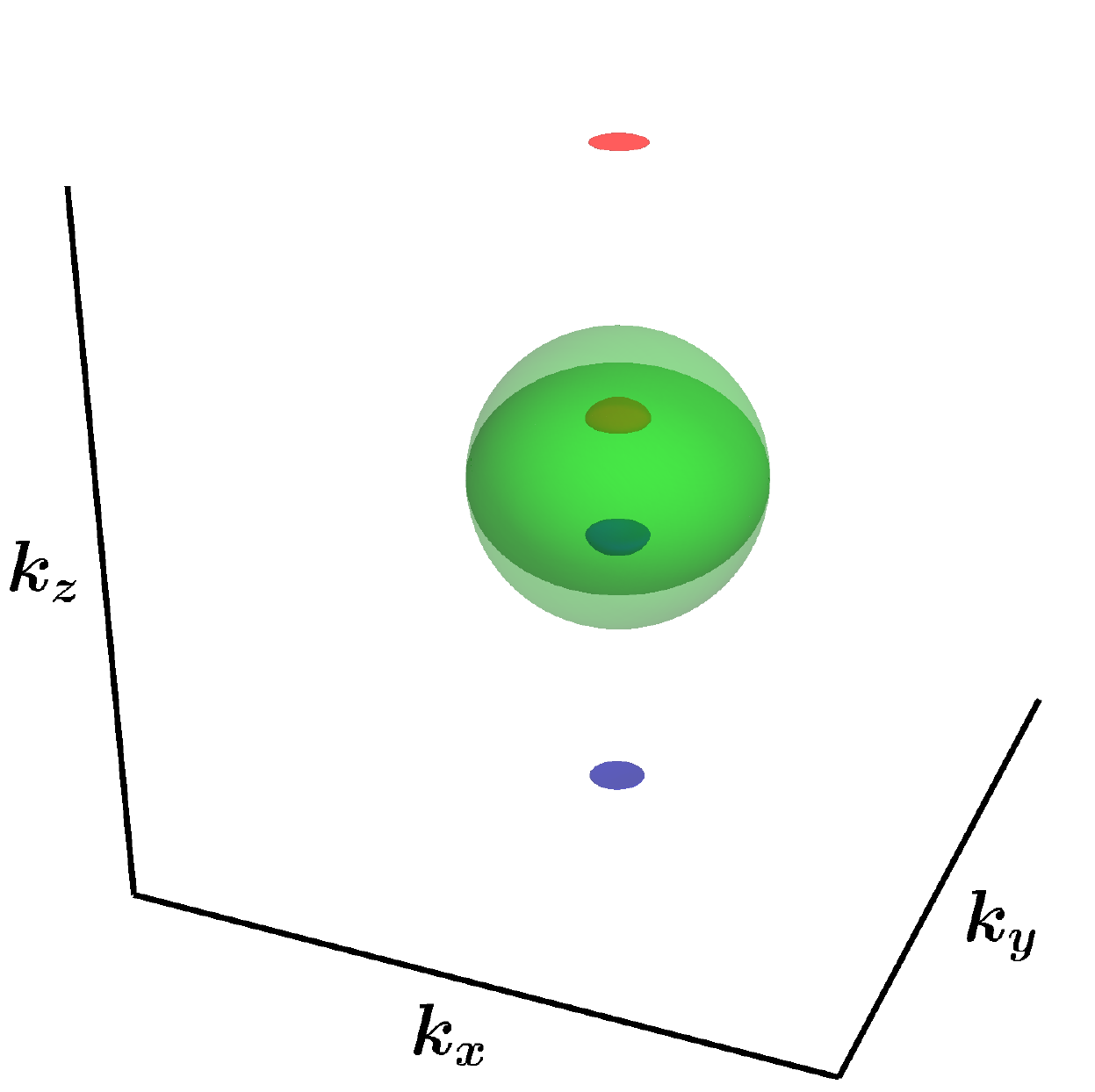}}
	    \end{minipage}
		\caption{(color online) Gap structure of the superconducting states : (a) $(s_z)$ state with $\mu\!<\!0$ $-$ A solid(dashed) ring indicates the nodal ring which is protected by the non-trivial winding number $1(-1)$. Each gapless region in $(s_z)$ state have two-fold degeneracy protected by the chiral symmetry (b) $(s_x\!+\!is_y)$ state with $\mu\!<\!0$, (c) $(s_x\!+\!is_y)$ state with $\mu\!=\!0$, and (d) $(s_x\!+\!is_y)$ state with $\mu\!>\!0$ $-$ Each surface indicates the non-degenerate Bogoliubov Fermi surface and the color represents its Chern number. Green, red, and blue indicates 0, -1, and 1 respectively. See main text for more details.}
		\label{fig:main}
\end{figure*}
For the time-reversal symmetric superconductor with $(s_z)$ pairing, $H_{BdG}(\bs{k})$ belongs to class BDI\cite{schnyder2008classification} and the spectrum is derived by the singular value decomposition of the following matrix, $\hat{h}(\boldsymbol{k})+i|\Delta|J_z\!=\!v(k_x,k_y,k_z+i|\Delta|) \cdot \boldsymbol{J} +(ck^2-\mu)\mathbb{I}_3$. The corresponding eigenvalues are given as,
\bea
\lambda_{s=1,0,-1}\!(\bs{k})\!=\!c|\bs{k}|^2 \!- \! \mu \!+ \! s\sqrt{v^2|  \bs{k}|^2 \!-\!|\Delta|^2 \!+\! 2i|\Delta|k_z}.
 \eea
For the momentum point $\bs{k}$ where $\lambda_{s}(\bs{k})\!=\!0$ is satisfied, the BdG energy spectrum become gapless. We find that $\lambda_{\pm1}(\bs{k})\!=\!0$ if $k_z\!=\!0$ and $c|\bs{k}|^2-\mu\!=\!\mp \sqrt{v^2|\bs{k}|^2-|\Delta|^2}$. These conditions define the two nodal lines when $\mu>c\Delta^2/v^2-v^2/4c$. Fig.\ref{fig:main} (a) show the two nodal rings, which are represented by solid and dashed line. The nodal rings are topologically protected by non-trivial winding number, $\omega\in\mathbb{Z}$, thus they are stable against any symmetry preserving perturbations. The winding number can be calculated as $\omega\!=\!\frac{1}{2\pi}\sum_s\int^{2\pi}_0 \partial_\theta \rm{arg}(\lambda_s)$, where the integration is taken along the loop that encircles each nodal line.\cite{qi2010topological,schnyder2008classification} We immediately find that the winding number of each solid and dashed nodal ring is $1$ and $-1$, respectively.
Similarly, the condition, $\lambda_{0}(\bs{k})\!=\!0$, defines the nodal surface, often referred to as Bogoliubov Fermi surface. Unlike the nodal lines, the Bogoliubov Fermi surface, characterized by $\lambda_{0}(\bs{k})\!=\!0$, is topologically trivial. This can be seen by including the additional odd-parity spin-singlet superconducting order parameter, which instantly gaps out the system. As a consequence, we expect topologically stable nodal lines for the time-reversal symmetric $(s_z)$ phase.

We now consider the gap structure of the time-reversal broken $(s_x\!+\!is_y)$ state. In this case, our system belongs to class D\cite{schnyder2008classification} and the gapless region can be calculated by finding $\bs{k}$ points which satisfy $\rm{det}$$[H_{BdG}(\bs{k})]\!=\!0$. This condition can be rewritten as,
\bea
4|\Delta|^2(|\Delta|^2+\bar{\mu}^2)(v^2k_z^2-\bar{\mu}^2)\!=\!\bar{\mu}^2(v^2|\bs{k}|^2-\bar{\mu}^2)^2
\eea
where $\bar{\mu}\!=\!ck^2-\mu$. The above single condition generally defines the surface in the three-dimensional momentum space. It realizes nodal surface in the BdG energy spectrum, which is now referred to as Bogoliubov Fermi surface (See Fig.\ref{fig:main} (b)-(d)). This Bogoliubov Fermi surface can be characterized by two distinct topological invariants.\cite{schnyder2008classification} The first is the $\mathbb{Z}_2$ valued number of occupied BdG bands. Each Bogoliubov Fermi surface is non-degenerate since the time-reversal symmetry is absent. This indicates that these surfaces are all topologically protected, since $\mathbb{Z}_2$ number always changes by 1 as the energy level cross the single Bogoliubov Fermi surface in the momentum space. The non-trivial $\mathbb{Z}_2$ number means that the Bogoliubov Fermi surface are locally stable until the two Bogoliubov Fermi surfaces pair-annihilate. In addition to the $\mathbb{Z}_2$ number, the Bogoliubov Fermi surface can be also characterized by non-trivial Chern number. In Fig.\ref{fig:main} (b)-(d), the Bogoliubov Fermi surface with the Chern number 1(-1) is colored blue(red). Rather than simply presenting the numerical results, we argue that the non-trivial Chern number is a necessary consequence of the well-known parity anomaly of two-dimensional Dirac fermion.\cite{fradkin1986physical,burkov2018quantum} First of all, we consider the adiabatic change from $(s_x)$ state to $(s_x\!+\!is_y)$ state by slightly turning on $(s_y)$ pairing. In $(s_x)$ state, $k_z\!=\!0$ plane can be viewed as nodal point superconductor with the two Dirac nodal points per each Bogoliubov Fermi surface. This Dirac nodal points are pinned at the zero energy states and they are the time-reversal partner to each other. As the infinitesimal time-reversal breaking $(s_y)$ pairing is turned on, the two Dirac points gaps out and must carry the Chern number $\pm 1/2$, which is analogous to the parity anomaly in the two-dimensional Dirac fermion. Since the time-reversal symmetry is broken, the effective mass gap of the Dirac points must be opposite with each other, therefore $k_z\!=\!0$ plane must be characterized by non-trivial Chern number. As a consequence, in full three-dimensional momentum space, each topologically protected nodal line is inflated into a couple of Bogoliubov Fermi surfaces possessing non-trivial Chern number $\pm 1$. In principle, the inflation of the nodal line to the nodal surface occur for each nodal line, and the total Chern number at $k_z\!=\!0$ plane can be canceled each other. However, each Bogoliubov Fermi surface must carry non-trivial Chern number until they pair-annihilate.

In conclusion, we have studied the triplet superconductivity of triple point fermions described by pseudospin-1 representation. In the superconductor composed of pseudospin-$1$ electrons, the even-parity paring can occur only with the spin-triplet pairing. 
Furthermore, we have shown that multiband interaction uniquely opens attractive triplet pairing channels. 
 Based on the Landau theory, we find two distinct triplet superconducting phases depending on the chemical potential $\mu$: the time-reversal symmetric $(s_z)$ state and the time-reversal broken $(s_x\!+\!is_y)$ state. 
In particular, $(s_x\!+\!is_y)$ phase is being favored when the chemical potential lies near or above the triple-band crossing point in such a way that middle band plays a role in electron pairing. Moreover, we find that the two states can be distinguished by different dimension of nodes and topological characteristics. 
Hence, we suggest the triplet superconductor is naturally stabilized in the triple point fermions in the presence of on-site interactions. In general, the superconducting instability is not limited to the on-site interactions, and therefore one may expect odd-parity superconductivity from the neighboring site pairings. In this case, we may expect $p$-wave spin-singlet and quintet states. The investigations on the possible odd-parity superconductivities would be an interesting topic for future study.

\begin{acknowledgments}
We thank Daniel Agterberg for valuable discussions. The authors acknowledge support from KAIST startup, BK21 plus program, KAIST and National Research Foundation Grant No. NRF2017R1A2B4008097. S.B.L thanks the hospitality at the Physics Department of University of California, San Diego. 
\end{acknowledgments}

\bibliography{J_1_super_bib}

\renewcommand{\thefigure}{S\arabic{figure}}
\setcounter{figure}{0}
\renewcommand{\theequation}{S\arabic{equation}}
\setcounter{equation}{0}

\begin{widetext}
\section{Supplementary Material for ``Triplet-Superconductivity in Triple-Band Crossings''}

\subsection{Ginzburg-Landau free energy and one-loop expansion}
\label{sec:glf}

In this section, we compute the coefficients of Ginzburg-Landau free energy $F(\vec{\Delta})$. We first introduce the propagator
\bea
G(K)=(ik_0 + (ck^2-\mu)\mathbb{I}_3+v\boldsymbol{k} \cdot \boldsymbol{J})^{-1}.
\eea
Here $K\equiv(k_0,\boldsymbol{k})$ and $k_0=2\pi(n+1/2)T$ denotes Matsubara frequency. Then, the free energy is written as,
\bea
F(\vec{\Delta})=\frac{1}{g} |\vec{\Delta}|^2 + T\sum_{m,n}\int_{\boldsymbol{k}}^{\Lambda} \frac{1}{2m} \text{tr}(-G(K)\Delta G(-K)^T\Delta^\dagger)^m,
\eea
where $\Delta=\sum_{a} J_a\gamma\Delta_a$. Let $F_{n}(\vec{\Delta})$ be the contribution to the free energy that contains $n$-th power of $\Delta_a$. We have
\bea
F_2(\vec{\Delta})&=&\frac{1}{g} |\vec{\Delta}|^2 -\frac{1}{2}\sum_{a,b}L_{ab}\Delta_{a}\Delta^*_{b},
\label{eq:f2}
\\
F_4(\vec{\Delta})&=&\frac{1}{4}\sum_{a,b,c,d}L_{abcd}\Delta_{a}\Delta^*_{b}\Delta_{c}\Delta^*_{d}
\label{eq:f4}
\eea
with
\bea
L_{ab}&=&T\sum_{k_0}\int_{\boldsymbol{k}}^{\Lambda} \text{tr}(G(K)J_a \gamma G(-K)^T (J_a\gamma)^\dagger),\\
L_{abcd}&=&T\sum_{k_0}\int_{\boldsymbol{k}}^{\Lambda} \text{tr}(G(K)J_a\gamma G(-K)^T(J_b\gamma)^\dagger G(K)J_c\gamma G(-K)^T(J_d\gamma)^\dagger).\\
\eea
Meanwhile, we can parametrize the general terms in $F_n(\vec{\Delta})$ accordingly.
\bea
F_2(\vec{\Delta})&=&r |\vec{\Delta}|^2,
\label{eq:free2}
\\
F_4(\vec{\Delta})&=&q_1 |\vec{\Delta}|^4 + q_2 \sum_{a}|\vec\Delta^* \mathcal{I}_a \vec\Delta|^2
\label{eq:free4}
\eea
where the matrix elements of $\mathcal{I}_a$ is given as $(\mathcal{I}_a)_{bc}=i\epsilon_{abc}$ and $\epsilon_{abc}$ is the Levi-Civita symbol.
Taking the specific configurations
\bea
\vec{\Delta}^1&=&(0,0,1),
\vec{\Delta}^2=\frac{1}{\sqrt{2}}(1,i,0)
\label{eq:ansatz}
\eea
we apply Eq.\ref{eq:f4} and
\bea
F_4(\vec{\Delta}^1)=q_1, 
F_4(\vec{\Delta}^2)=q_1+q_2
\eea
to get the coefficients $q_1$ and $q_2$. With introducing $\hat{k}_0=k_0/T, \hat{\boldsymbol{k}}=\boldsymbol{k}/\sqrt{T}$, $\hat{\mu}=\mu/T$, $\hat{v}=v/\sqrt{T}$, and $\hat\Lambda=\hat\mu/\hat v$ we find
\bea
q_1&=&\frac{1}{T^{3/2}}\int_{\hat{\boldsymbol{k}}}^{\hat{\Lambda}}\sum_n\frac{g_1}{\left((\hat{\mu} -c \hat{k}^2)^2+\hat{k}_0^2\right)^2 \left((\hat{k} (\hat{v}-c \hat{k})+\hat{\mu} )^2+\hat{k}_0^2\right)^2 \left((\hat{\mu} -\hat{k} (c \hat{k}+\hat{v}))^2+\hat{k}_0^2\right)^2},
\label{eq:q1}
\\
q_2&=&\frac{1}{T^{3/2}}\int_{\hat{\boldsymbol{k}}}^{\hat{\Lambda}}\sum_n\frac{g_2}{\left((\hat{\mu} -c \hat{k}^2)^2+\hat{k}_0^2\right)^2 \left((\hat{k} (\hat{v}-c \hat{k})+\hat{\mu} )^2+\hat{k}_0^2\right)^2 \left((\hat{\mu} -\hat{k} (c \hat{k}+\hat{v}))^2+\hat{k}_0^2\right)^2}
\label{eq:q2}
\eea
where
\bea
\nonumber
g_1&=&\frac{2}{15} \pi\Big(\left(\hat{\mu} -c \hat k^2\right)^4 \left(15 c^4 \hat k^8+10 c^2 \hat k^6 \hat{v}^2+10 \hat k^2 \hat{\mu} ^2 \left(9 c^2 \hat k^2+\hat{v}^2\right)-20 c \hat k^4 \hat{\mu}  \left(3 c^2 \hat k^2+\hat{v}^2\right)-60 c \hat k^2 \hat{\mu} ^3-\hat k^4 \hat{v}^4+15 \hat{\mu} ^4\right)\\
\nonumber
&+&5 \hat k_0^2\Big(2 \left(\hat{\mu} -c \hat k^2\right)^2 \left(6 c^4 \hat k^8-24 c^3 \hat k^6 \hat{\mu} +c^2 \left(5 \hat k^6 \hat{v}^2+36 \hat k^4 \hat{\mu} ^2\right)-2 c \left(5 \hat k^4 \hat{\mu}  \hat{v}^2+12 \hat k^2 \hat{\mu} ^3\right)-5 \hat k^4 \hat{v}^4+5 \hat k^2 \hat{\mu} ^2 \hat{v}^2+6 \hat{\mu} ^4\right)\\
\nonumber
&+&\hat k_0^2\Big(18 c^4 \hat k^8-72 c^3 \hat k^6 \hat{\mu} +2 c^2 \left(7 \hat k^6 \hat{v}^2+54 \hat k^4 \hat{\mu} ^2\right)-4 c \left(7 \hat k^4 \hat{\mu}  \hat{v}^2+18 \hat k^2 \hat{\mu} ^3\right)+3 \hat k^4 \hat{v}^4+14 \hat k^2 \hat{\mu} ^2 \hat{v}^2+18 \hat{\mu} ^4\\
&+&3 \hat k_0^2 \left(4 c^2 \hat k^4-8 c \hat k^2 \hat{\mu} +2 \hat k^2 \hat{v}^2+\hat k_0^2+4 \hat{\mu} ^2\right)\Big)\Big)\Big),
\label{eq:g1}
\\
\nonumber
g_2&=&\frac{2}{15} \pi  \hat k^2 \hat{v}^2\Big(\left(\hat{\mu} -c \hat k^2\right)^2 \left(5 c^4 \hat k^8-20 c^3 \hat k^6 \hat{\mu} +6 c^2 \hat k^4 \left(5 \hat{\mu} ^2-3 \hat k^2 \hat{v}^2\right)+4 c \left(9 \hat k^4 \hat{\mu}  \hat{v}^2-5 \hat k^2 \hat{\mu} ^3\right)+5 \hat k^4 \hat{v}^4-18 \hat k^2 \hat{\mu} ^2 \hat{v}^2+5 \hat{\mu} ^4\right)\\
\nonumber
&+&5\hat k_0^2\Big(c^4 \hat k^8-4 c^3 \hat k^6 \hat{\mu} +c^2 \left(6 \hat k^4 \hat{\mu} ^2-4 \hat k^6 \hat{v}^2\right)+c \left(8 \hat k^4 \hat{\mu}  \hat{v}^2-4 \hat k^2 \hat{\mu} ^3\right)-\hat k^4 \hat{v}^4-4 \hat k^2 \hat{\mu} ^2 \hat{v}^2+\hat{\mu} ^4\\
&-&\hat k_0^2 \left(c^2 \hat k^4-2 c \hat k^2 \hat{\mu} +2 \hat k^2 \hat{v}^2+\hat k_0^2+\hat{\mu} ^2\right)\Big)\Big).
\label{eq:g2}
\eea
Utilizing Eq.\ref{eq:q2} and Eq.\ref{eq:g2}, we investigate the sign of $q_2$ varying $\hat \mu$ and $\hat v$ while keeping $c=1/10$. Then we acquire the phase diagram as given in the main text.

In the limit where the middle band is perfectly flat($c=0$), $q_1$ and $q_2$ can be simplified as below with normalizing the field, $\psi$, such that $\hat{v}$ becomes unity.
\bea
q_1&=&\frac{1}{T^{3/2}}\frac{2\pi}{15} \int_{\hat{\boldsymbol{k}}}^{\hat{\mu}}\sum_n\frac{\hat k^4 \left(-50 \hat k_0^2 \hat \mu ^2+15 \hat k_0^4-\hat \mu ^4\right)+10 \hat k^2 \left(\hat k_0^2+\hat \mu ^2\right)^2 \left(3 \hat k_0^2+\hat \mu ^2\right)+15 \left(\hat k_0^2+\hat \mu ^2\right)^4}{\left(\hat k_0^2+\hat \mu ^2\right)^2 \left((\hat \mu -\hat k)^2+\hat k_0^2\right)^2 \left((\hat k+\hat \mu )^2+\hat k_0^2\right)^2},
\label{eq:q1f}
\\
q_2&=&\frac{1}{T^{3/2}}\frac{2\pi}{15} \int_{\hat{\boldsymbol{k}}}^{\hat{\mu}}\sum_n\frac{ \hat k^2 \left(5 \hat k^4 \hat \mu ^2-18 \hat k^2 \hat \mu ^4-5 \hat k_0^2 \left(\hat k^4+4 \hat k^2 \hat \mu ^2+\hat k_0^2 \left(2 \hat k^2+\hat \mu ^2\right)+\hat k_0^4-\hat \mu ^4\right)+5 \hat \mu ^6\right)}{\left(\hat k_0^2+\hat \mu ^2\right)^2 \left((\hat \mu -\hat k)^2+\hat k_0^2\right)^2 \left((\hat k+\hat \mu )^2+\hat k_0^2\right)^2}
\label{eq:q2f}
\eea
In Fig.\ref{fig:fcp} we display the plot of $q_1T^{3/2}$ and $-q_2T^{3/2}$ using Eq.\ref{eq:q1f} and Eq.\ref{eq:q2f}. We observe that the free energy is stable ($q_1T^{3/2}>0$) and time reversal broken phase is energetically favored ($q_2T^{3/2}<0$) when the middle band is perfectly flat ($c=0$).

\begin{figure}[h]
	\includegraphics[width=0.6\columnwidth]{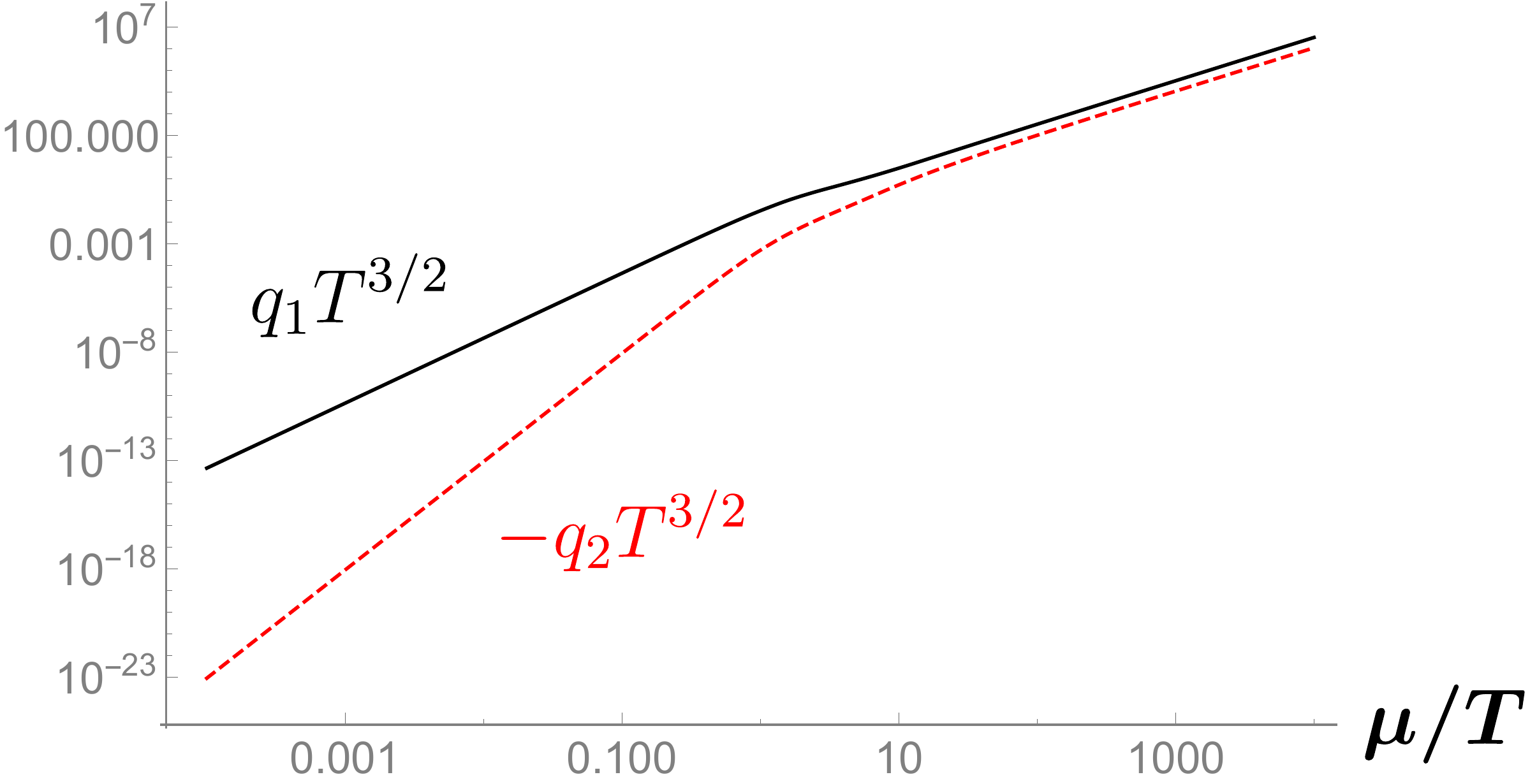}
	\caption{(color online) Plot of $q_1T^{3/2}$ (solid black line) and $-q_2T^{3/2}$ (dashed red line) within one-loop calculation.
	}
	\label{fig:fcp}
\end{figure}

\end{widetext}

\end{document}